# Creating a Mixed-Reality Installation with Families through Theatrical Co-Design

Pavlos Panagiotidis

Mixed Reality Lab, School of Computer Science, University of Nottingham, Nottingham, United Kingdom, pavlos.panagiotidis@nottingham.ac.uk

Roma Patel

Makers of Imaginary Worlds, Nottingham, United Kingdom, roma@makersofimaginaryworlds.co.uk

Boriana Koleva

Mixed Reality Lab, School of Computer Science, University of Nottingham, Nottingham, United Kingdom, b.koleva@nottingham.ac.uk

Steve Benford

Mixed Reality Lab, School of Computer Science, University of Nottingham, Nottingham, United Kingdom, steve.benford@nottingham.ac.uk

Jocelyn Spence

Mixed Reality Lab, School of Computer Science, University of Nottingham, Nottingham, United Kingdom, jocelyn.spence4@nottingham.ac.uk

Paul Tennent

Mixed Reality Lab, School of Computer Science, University of Nottingham, Nottingham, United Kingdom, paul.tennent@nottingham.ac.uk

Juan Pablo Martinez Avila

Mixed Reality Lab, School of Computer Science, University of Nottingham, Nottingham, United Kingdom, j.avila@nottingham.ac.uk

Laurence Cliffe

Mixed Reality Lab, School of Computer Science, University of Nottingham, Nottingham, United Kingdom, laurence.cliffe@nottingham.ac.uk

Co-designing with families for environmental sustainability relies on participatory imagination, yet habitual family roles and uneven participation, especially between adults and young children, often constrain it. A second challenge is continuity: workshop relationships and embodied ways of working do not easily survive into the final design, where artefacts travel more readily than roles or interactional dynamics. We report on a nationally toured mixed-reality installation developed through applied-theatre-led co-design with families. Across three workshops and user testing, applied theatre methods supported families to co-create narratives, artefacts, and interactional roles that shaped the public event. We show how theatrical co-design can rebalance child–adult participation through playful status shifts, and how selected workshop dynamics can be re-staged within a public mixed-reality installation. We contribute a theatrical account of participatory design in which designers work not only with artefacts and ideas, but with roles, rhythms, authority, and the social conditions that support collective imagination.

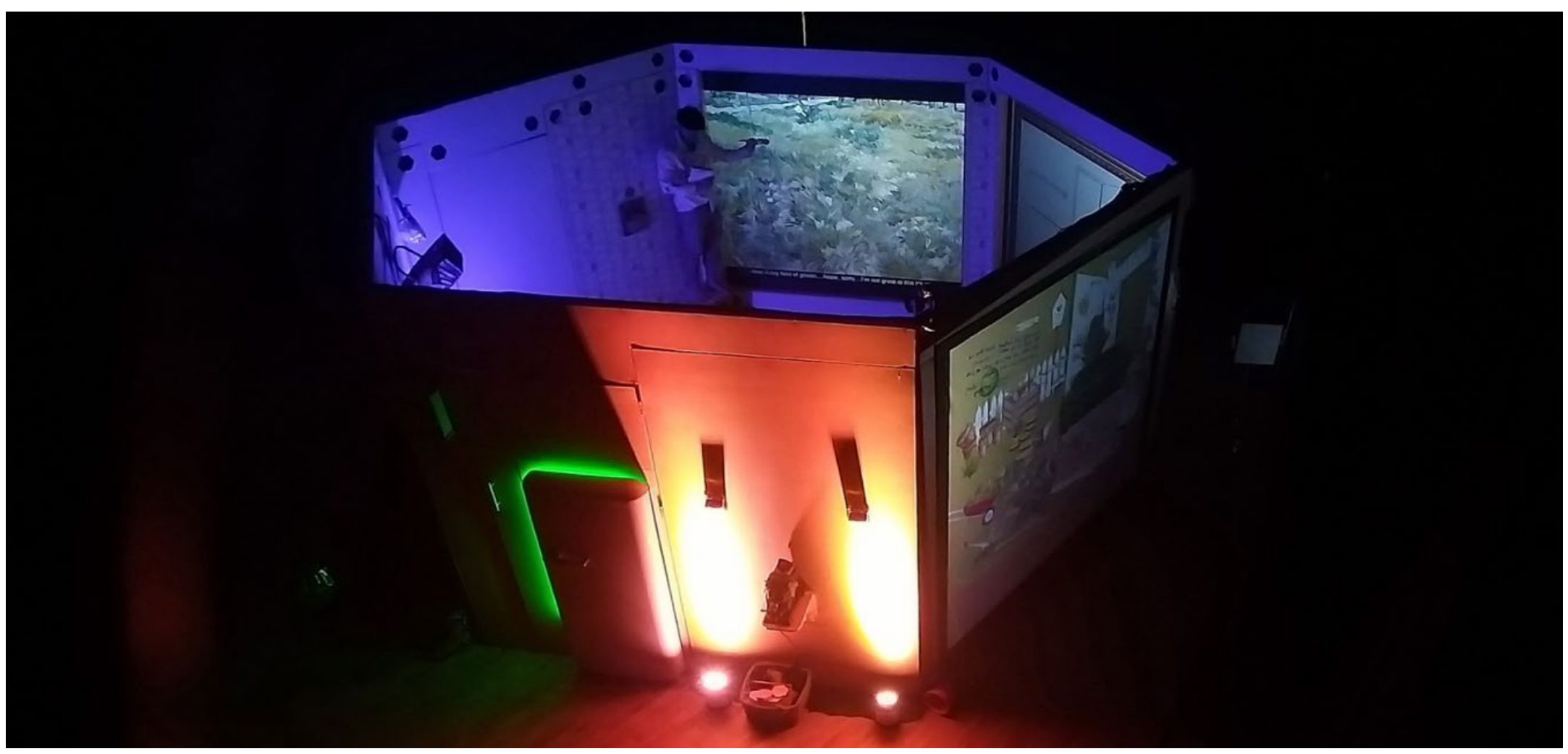

Figure 1. *HOME: Zero* was co-created with families using theatrical methods to engage audiences in discussing and pledging to environmental interventions in their own family lives.

## 1 INTRODUCTION

We report on *HOME: Zero*, a performance-led, tourable mixed-reality artwork co-designed with families. The work was created by Makers of Imaginary Worlds, commissioned by Nesta and National Gallery X (NGX), and shown across national and community venues. We use this case to examine how applied-theatre co-design can support more equitable intergenerational participation in an environmental sustainability context, and how theatrical qualities can be carried into a public-facing event.

Designing technologies that engage people with environmental sustainability is widely recognised as a wicked problem: complex, open-ended, and resistant to straightforward solutions [6, 13, 20]. Within HCI, ecologically engaged artworks have been used to prompt reflection on environmental issues [7, 10, 17]. Co-design is well suited to such contexts because it supports collaboration and broader idea generation around complex societal problems [5, 19]. In our case, families were engaged as experts on their own homes, with children and adults both bringing relevant perspectives on domestic practices and climate-related change. However, family co-design raises challenges. Earlier work has shown the value of cultural probes, low-tech prototyping, and family-centred design activities [14, 18, 27, 28]. Yet these studies also highlight the difficulty of balancing children's and adults' contributions and ensuring that participants feel valued regardless of age or experience. Yip et al. further show that intergenerational co-design is slow and transitional, with parents and children moving unevenly toward equal design partnership [29, 30]. This matters in sustainability contexts, where the aim is not only to generate ideas, but to support families in collectively imagining and committing to change.

Applied theatre offers a relevant set of practices for this problem. It has a long tradition of involving non-professional participants in collective reflection and social change, including Boal's approach of turning audiences into active "spect-actors" [3, 4]. It also provides methods for play, embodiment, role-taking, and status negotiation [11, 12]. In HCI, theatre-based methods have supported embodied co-creation, emotional reflection, and inclusive participation [16, 21, 23–25]. Yet much of this work remains researcher-led or pedagogical, with children often participating separately from adults and outcomes usually remaining within workshops, prototypes, or classroom settings. What remains less explored is how applied theatre can structure intergenerational family co-design under professional production constraints, and how the resulting roles and participatory conditions can move from workshops into a public mixed-reality artwork.

This paper analyses the theatrical co-design process behind *HOME: Zero*, examining how applied theatre methods structured family participation and shaped the transition from workshop activity to public mixed-reality encounter. We distil two design orientations for family co-design around complex social and environmental issues: rebalancing child–adult power through playful status shifts, and re-staging selected workshop dynamics within the public mixed-reality installation.

## 2 METHOD

We frame the project as Performance-led Research in the Wild [2]: a professional production developed under real commissioning constraints and analysed retrospectively as an HCI design case. Our corpus comprised materials from across the pitch-to-tour process, including workshop plans and notes, family-generated drawings and prototypes, photographic documentation, user-testing materials, design-change records, installation documentation, and a semi-structured interview with the performer–facilitator. We analysed these materials through situated design reflection: building a timeline linking workshop activities to later design decisions, identifying critical moments of participation and workshop-to-installation transfer, and synthesising these into the two design orientations discussed below. The lead practitioner/artist is a co-author, and their account was treated as reflexive memoing, checked against project documents, artefacts, and the facilitator interview. Institutional ethics approval was obtained, and consent was gathered from participating families, including child-facing information sheets.

## 3 THE INSTALLATION

Before turning to the co-design process, we briefly outline *HOME: Zero* and the audience journey illustrated in Figure 2. The installation was a large hexagonal mixed-reality room, entered through a recycled 1950s-style fridge door. Inside, families encountered projection walls, speakers, a central "upcycler" creating a Pepper's Ghost hologram effect, and several physical interaction props, including a salad spinner, Victorian bellows, and a bespoke wooden controller. The experience lasted around twenty-five minutes and welcomed six to eight participants at a time. Families were greeted by a performer who used an "imagination scanner" to playfully measure imagination levels, always telling children theirs were higher than adults, before leading everyone inside. Once inside, families were introduced to Kitty, the disembodied voice of the homeowner.

The narrative began with a projected door opening onto a devastated world outside. Kitty asked participants to pause time and search for solutions. The projected walls then revealed paintings from NGX's collection, alongside a canvas sketch of a room and a sketchbook of notes and hints. Families used the salad spinner to scroll through paintings, the bellows to select objects, and the upcycler to capture them as floating holograms. They then used the wooden controller to transform these objects and place them onto a blank canvas as possible environmental solutions. At the end, each family received a small hexagonal block on which to write an environmental pledge. These pledges were added to the exterior wall of the installation, forming a growing artwork of shared commitments.

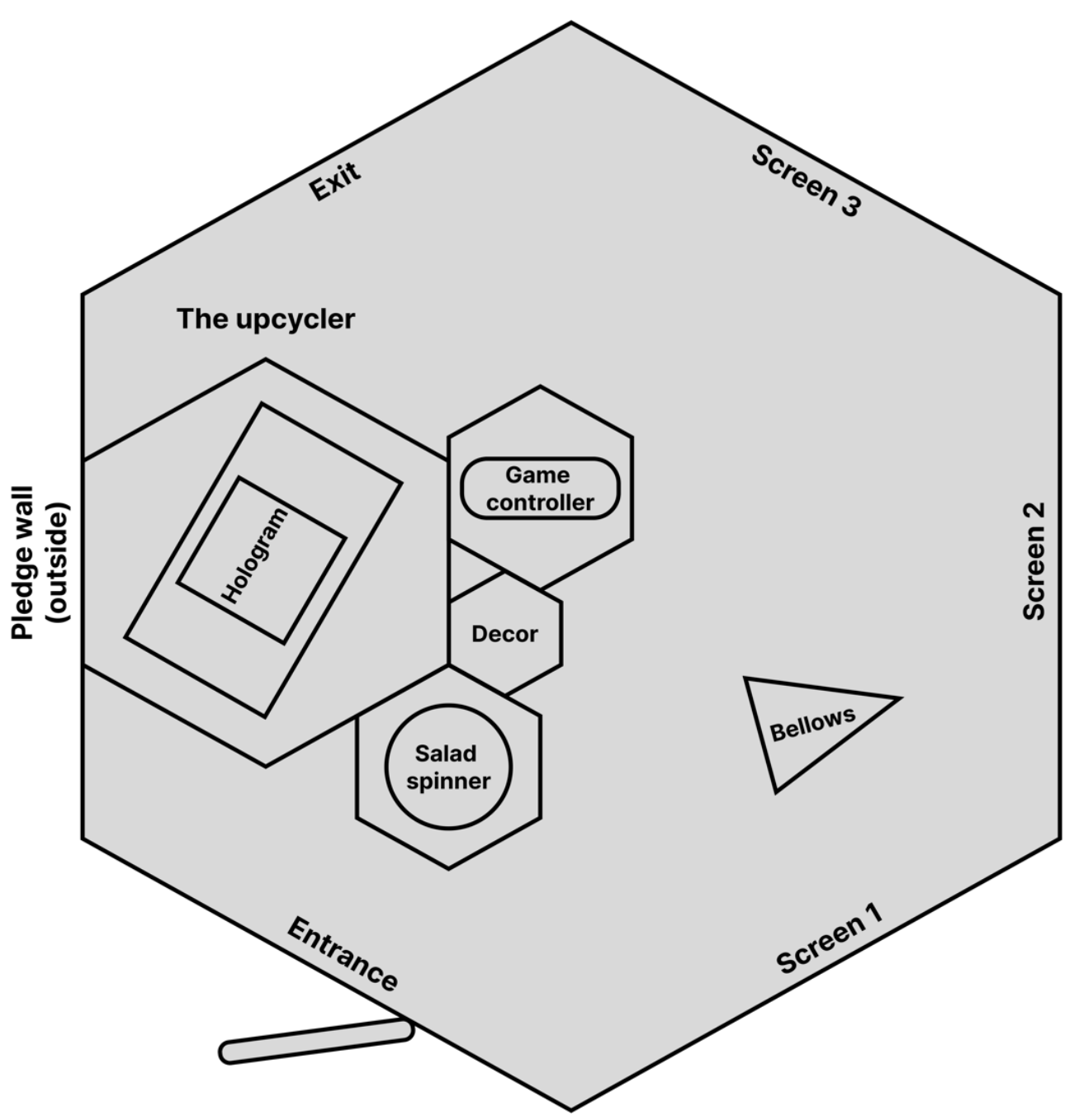

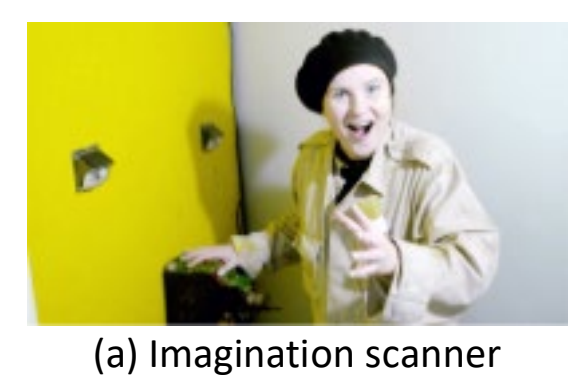
(a) Imagination scanner

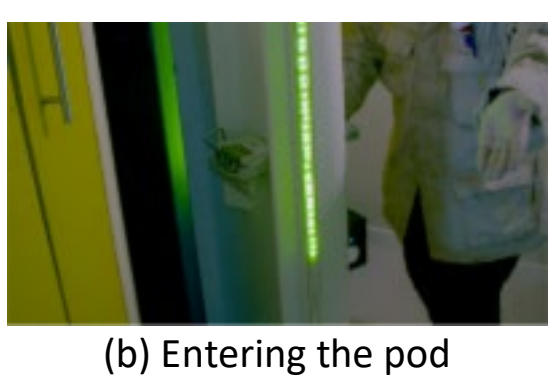
(b) Entering the pod

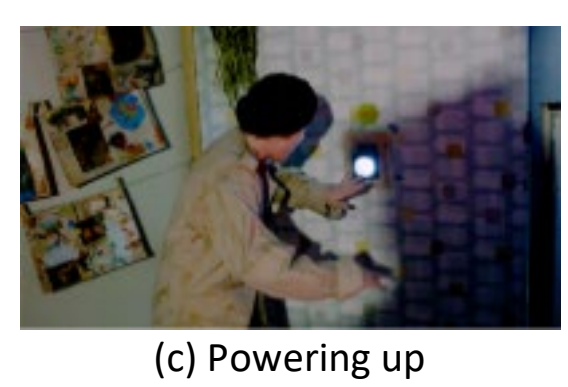
(c) Powering up

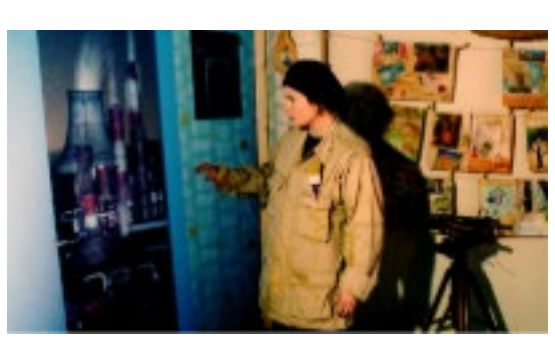
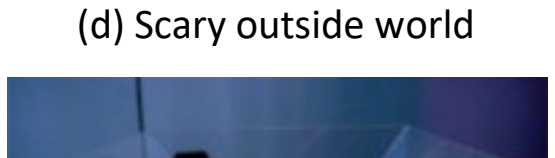
(d) Scary outside world

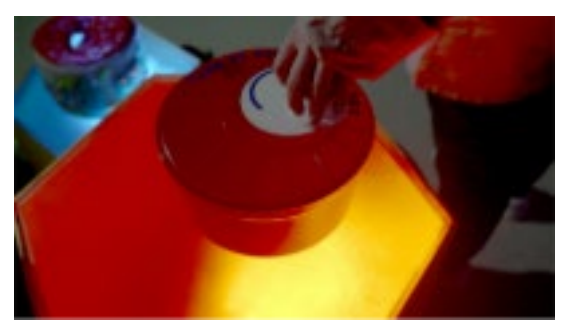

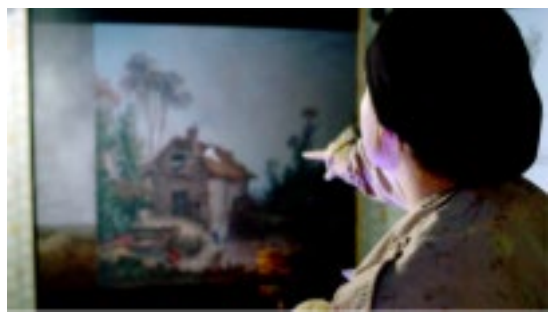

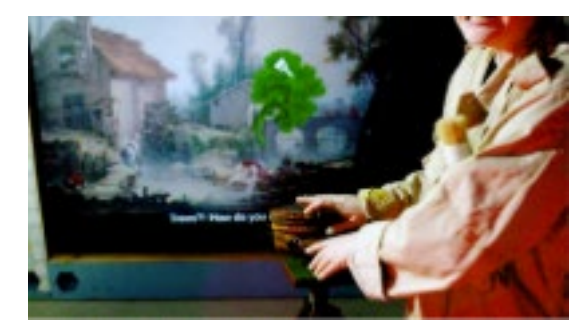

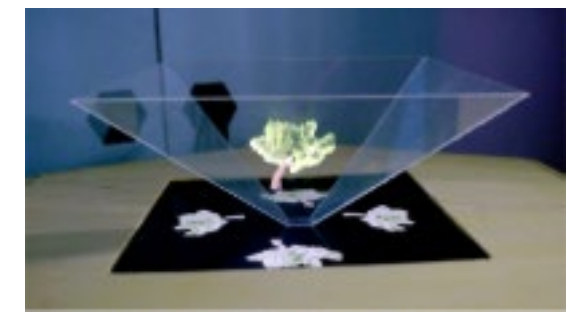

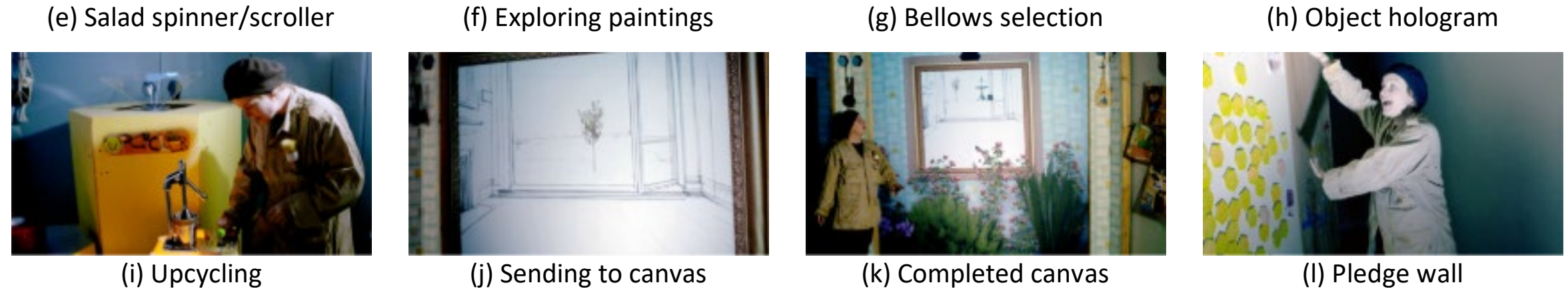

Figure 2. Overview of the installation - top: spatial map of the pod; bottom: storyboard of the audience journey (performer image used with permission).

## 4 CO-DESIGNING PROCESS

The creative team's approach was to place children at the centre of the experience, while enabling families to participate as a collective unit. Co-design was built into the project from the original pitch to National Gallery X's open call, rather than added later as a consultation stage. A key decision was to integrate applied theatre practices into the process by engaging a professional performer/facilitator to lead the workshops. Over seven months, from the initial pitch in November 2021 to the public exhibition in May 2022, the installation developed through three co-design workshops and a user-testing session.

Local families with children aged 6–10 were recruited through an online campaign and direct mailing by Lakeside Arts. Seventeen families registered, eleven attended the first workshop, and nine stayed with the project through the full process. Alongside the families, the workshops involved the lead artist, who set the aesthetic direction and curated emerging ideas, and the performer–facilitator, who led the theatrical activities. Design decisions were partly collective, such as the choice of the kitchen as the central domestic setting, and partly curated, with recurring, feasible, or dramaturgically useful ideas taken forward. This meant that family contributions remained visible in the final installation, including the bellows mechanism, which developed from a child's "sucking machine," and the "stop time" feature, which emerged from workshop discussions.

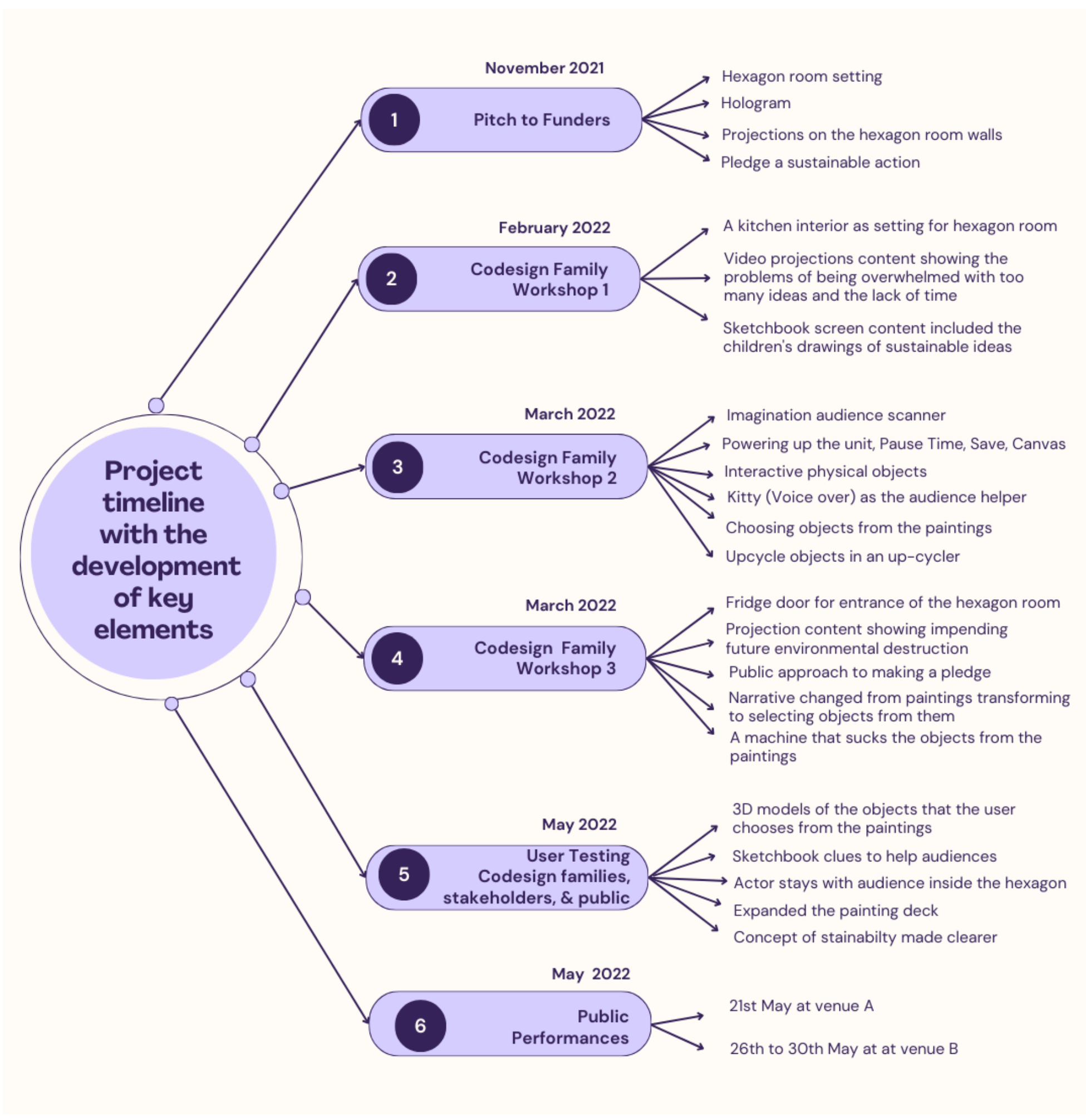

Figure 3. Timeline of the main project phases and development of key elements of the experience

Table 1: Workshop Participants

| Family | Number of Adults | Number of Children | Children age(s) |
|---|---|---|---|
| Family of 4 | 2 | 2 | 7, 9 |
| Family of 3 | 2 | 1 | 7 |
| Family of 4 | 2 | 2 | 6, 10 |
| Family of 3 | 1 | 2 | 9 (Twins) |
| Family of 2 | 1 | 1 | 10 |
| Family of 2 | 1 | 1 | 6 |
| Family of 3 | 1 | 2 | 6, 8 |
| Family of 3 | 2 | 1 | 8 |
| Family of 2 | 1 | 1 | 8 |

### 4.1 Workshops and the Role of the Facilitator

The first workshop aimed to understand how families interpreted the selected paintings and to identify which room in the home should frame the experience. Families first looked at the paintings and wrote or drew what these made them feel or think. They were then asked which room in their house they could make "perfect for the planet" and to draw or write possible ideas. Finally, each family performed their solutions to the rest of the group using only their bodies. This last activity was important because it moved participation away from verbal reasoning, where adults often hold more authority, and drew on physical theatre methods that allow children to express ideas through gesture, movement, and spatial interaction [21].

Some paintings were found to be difficult to connect to sustainability, so the collection was revised to better support themes such as biodiversity, water conservation, sustainable energy, repair culture, greener homes, heat conservation, rewilding, and growing and composting. The kitchen emerged as the most popular room, and this became the setting for the installation. Families' drawings also became material for the final Sketchbook screen, where they appeared as clues and examples of "good ideas" for visitors. At the same time, the final discussion revealed a participation problem. Children showed considerable awareness of environmental issues and contributed actively during embodied tasks, but the reflective discussion was still largely dominated by adult voices. This shaped the design of the second workshop, where more explicit steps were taken to position children as creative leaders.

The second workshop focused on how audiences might interact with the installation. As families entered, the facilitator used an "imagination scanner" to measure each participant's level of imagination. The children were deliberately given the highest scores and received special badges. This playful device was used to empower children to take the lead and to encourage adults to become more playful and supportive. The workshop then moved into a physical ice-breaker, Cops and Robbers, which asked families to chase, mime, and imagine together. This helped establish a shared playful mode before the design work began.

Families then considered how a painting might change in relation to sustainability and prototyped possible interaction mechanisms for the installation. To help them imagine the future space, the team marked out the hexagonal room on the floor and explained that animated projections could appear on the walls. Families were invited to imagine a central table or control panel that visitors could use to stop the deterioration of the paintings. They created prototypes from craft materials and found objects, then presented them bodily inside the marked-out installation space. This was followed by a discussion about similarities and differences across designs, and about how to ensure equal voices between children and adults.

The outputs from this workshop substantially changed the interaction design. The original idea of a projected tabletop was replaced by a set of physical controls distributed around the room. Participant designs included turning wheels, powering mechanisms, and using multiple objects rather than a single interface. This led to the repurposed salad spinner, Victorian bellows, and bespoke wooden controller used in the final upcycling interaction. The idea that people should work together from different positions in the room also reinforced teamwork and equal participation as design values. Key interactions emerged here too, including choosing objects from paintings and transforming them into environmental solutions. Families also expressed interest in receiving help or guidance, which contributed to the development of Kitty, the disembodied voice of the homeowner. In the final discussion, adults commented that the imagination scanner reminded them not to constrain children's ideas and helped everyone enter a more creative mode. It was therefore retained as the first interaction in the public installation.

The third workshop focused on narrative development and pledges. After another physical warm-up, families explored an interactive story structure using Twine, making collective choices about how the narrative might unfold. The task invited improvised responses to a situation in which paintings in the gallery changed unpredictably and participants had to act together to stop this. This helped shift the narrative from paintings simply transforming to visitors actively selecting objects from paintings and using a machine to "suck" and upcycle them, while the projected walls showed impending environmental destruction. This consolidated ideas already emerging in Workshop 2 around choosing objects, upcycling them, and using them to repair or reimagine a domestic environment. The second part of the workshop focused on pledges. Families created commitments for positive change at home, discussed possible barriers, and presented their pledges publicly. This process evolved into the pledge wall, where audience families wrote environmental commitments on small hexagonal blocks and added them to the outside of the installation. The pledge activity was important because it extended the installation beyond imaginative play alone: it gave families a public, tangible way to connect the experience to possible action after leaving the event.

The performer/facilitator's role also became increasingly important across the process. In the workshops, he used his performance background to model safe engagement in playfulness and silliness, helping adults and children relax into a more imaginative mode. He described this less as showing participants what to do and more as showing that it was "safe to" participate playfully. This mattered in a family setting because it softened familiar adult–child roles and supported situations where children could be treated as imaginative experts, while adults were encouraged to listen and build from their ideas. By user testing, this facilitative role had become part of the installation itself: the performer remained inside the hexagon with visitors, guiding attention, encouraging teamwork, and helping sustain the participatory conditions developed in the workshops. This reinforced the sense that the co-design process was not only a source of material for the installation, but also a rehearsal of the social dynamics that the final event needed to support.

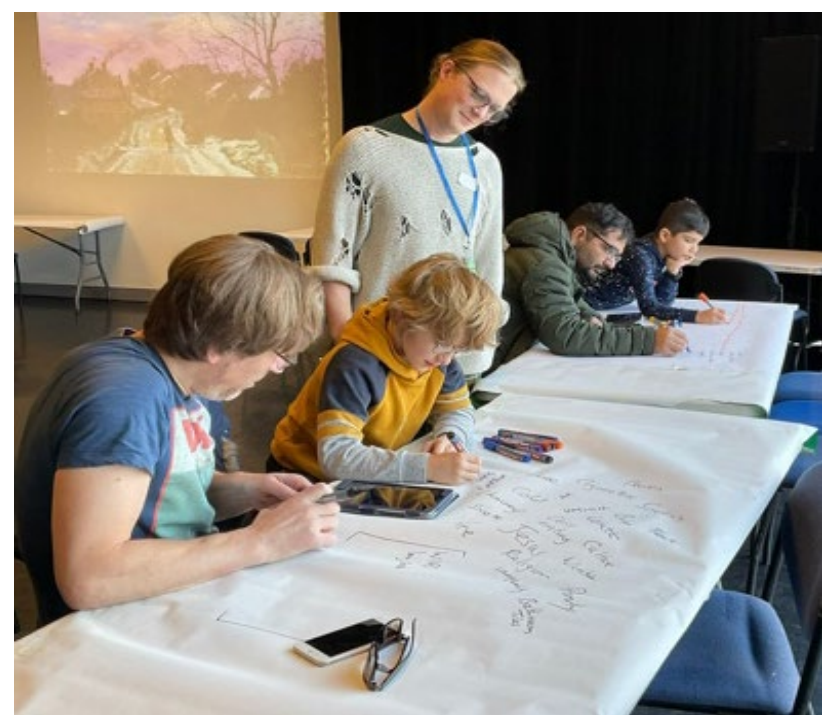

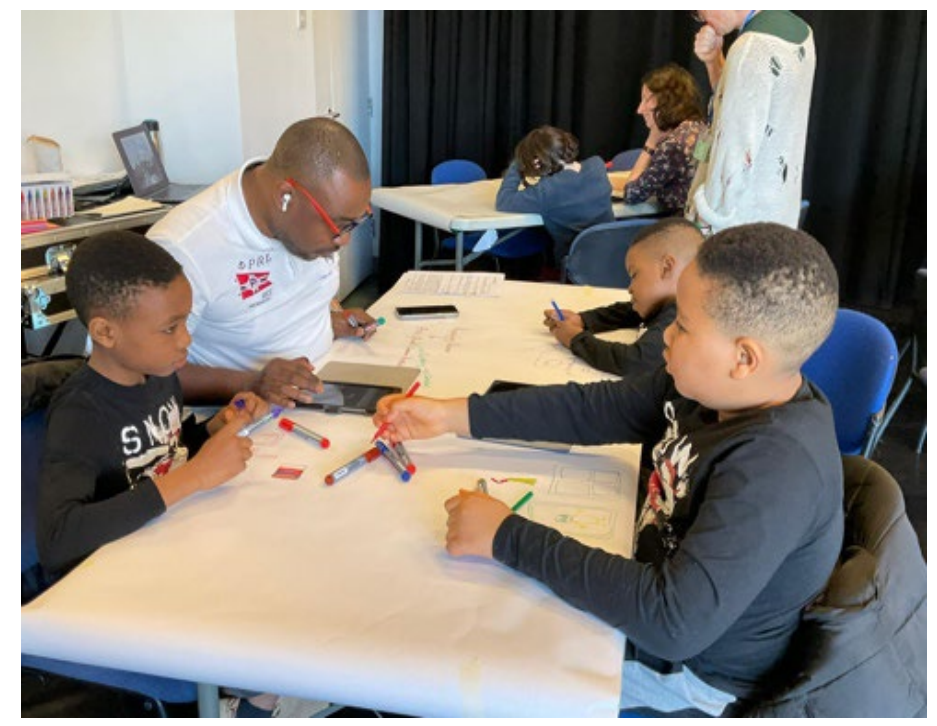

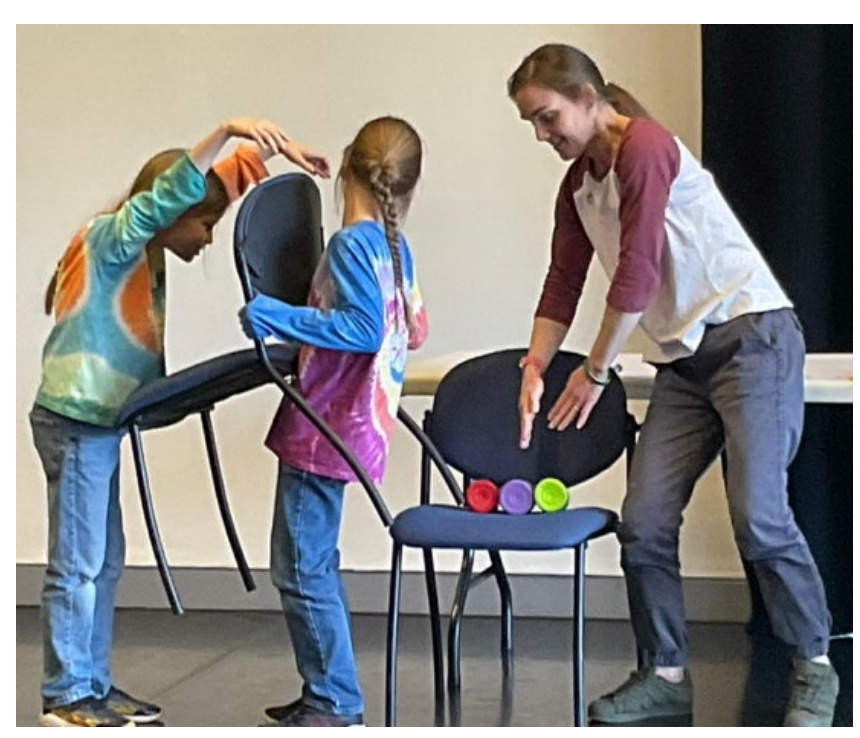

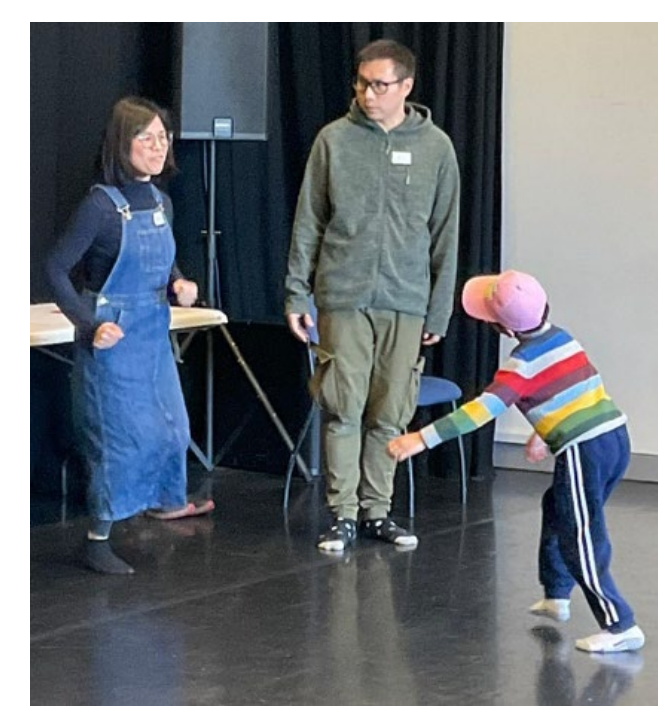

Figure 4. Families drawing (top) and performing (bottom) sustainable ideas

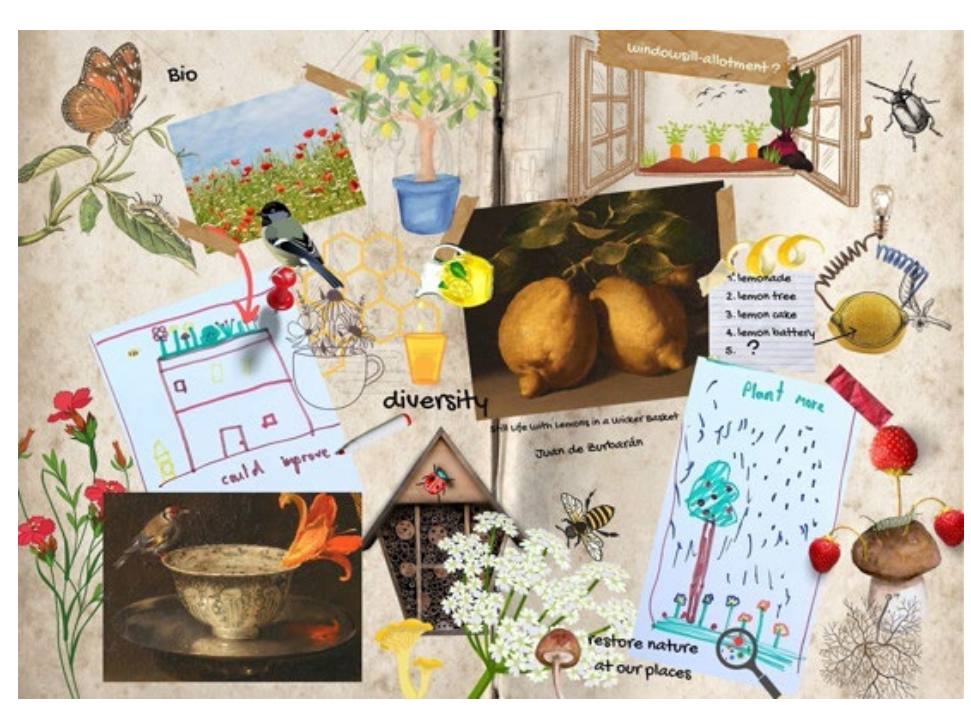

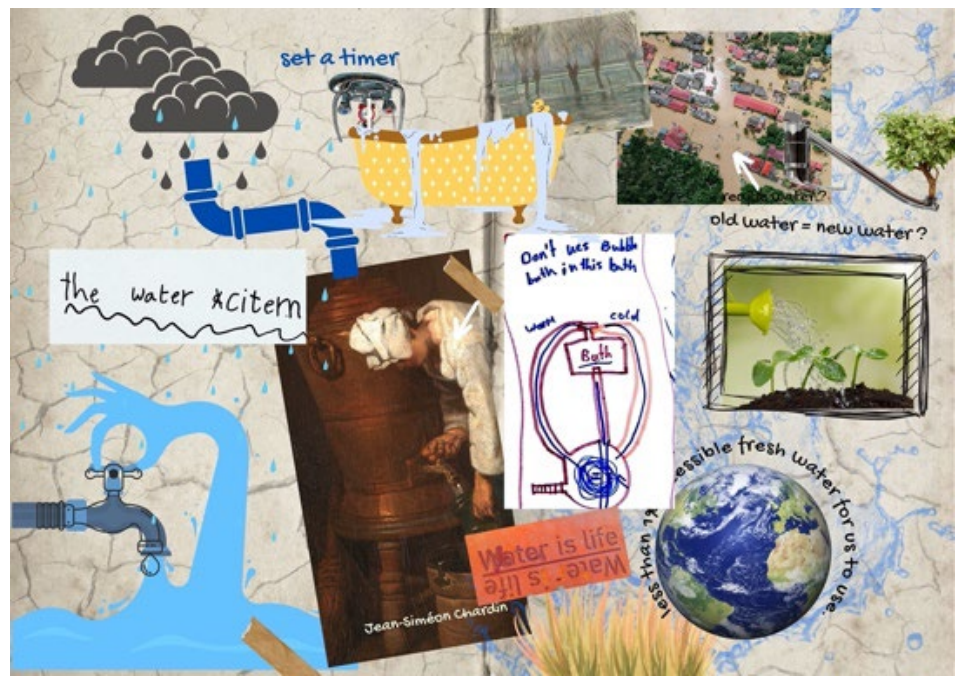

Figure 5. Examples of drawings integrated into the sketchbook display of the installation

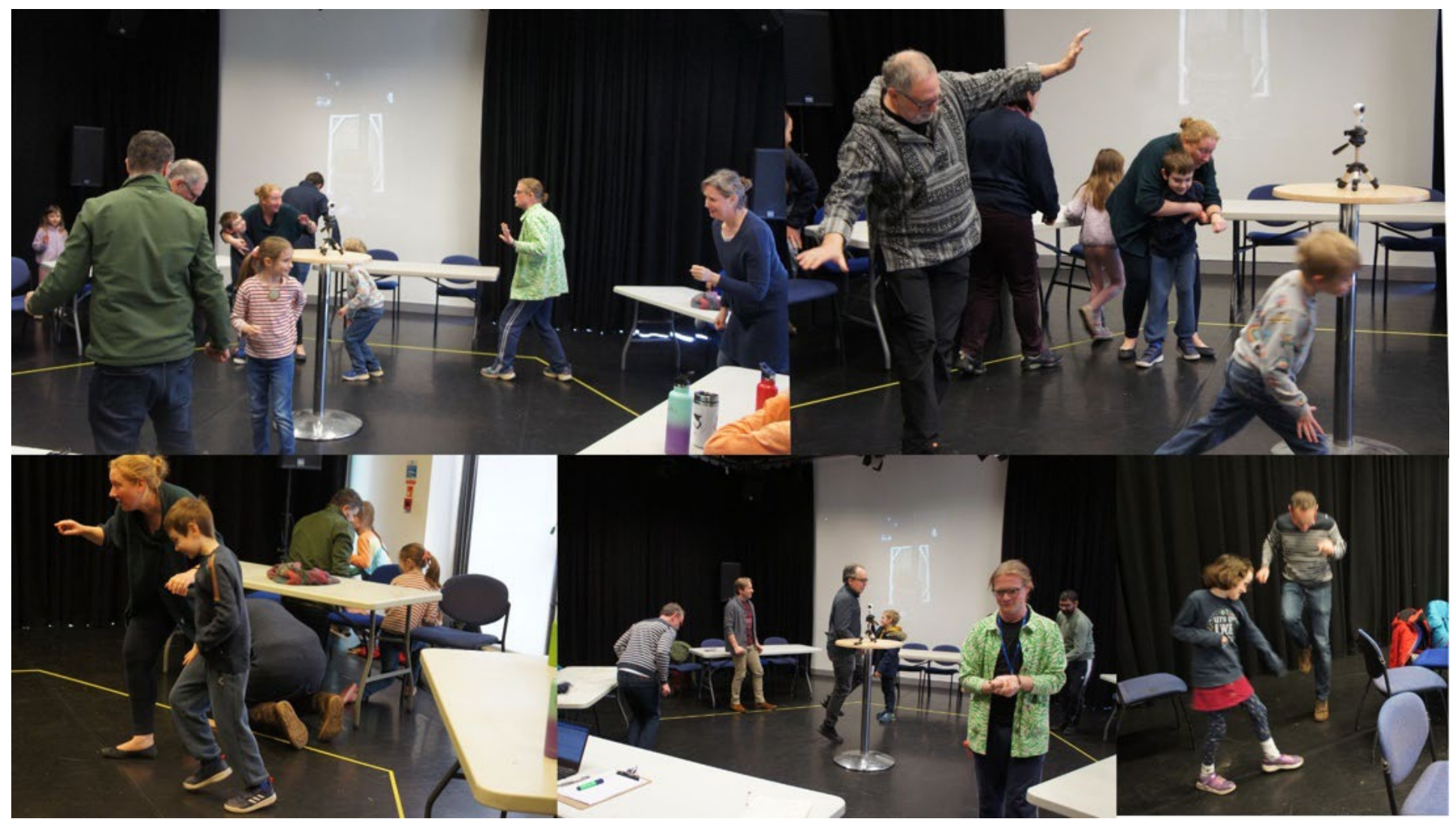

Figure 6. Cops and Robbers ice breaker game

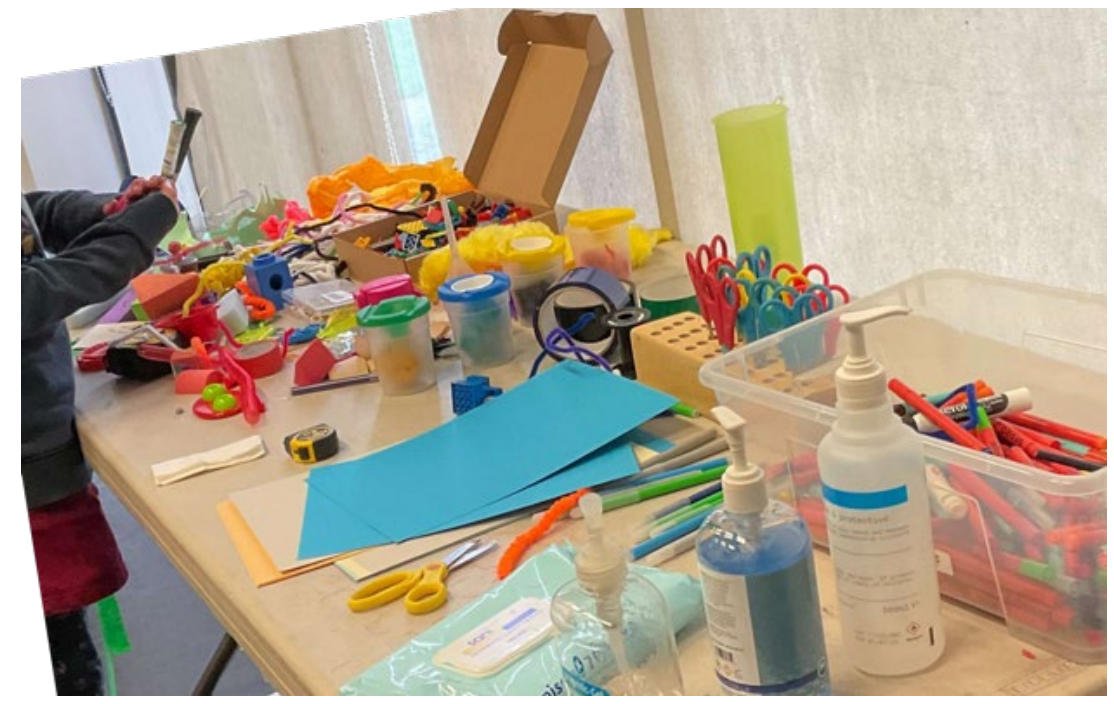

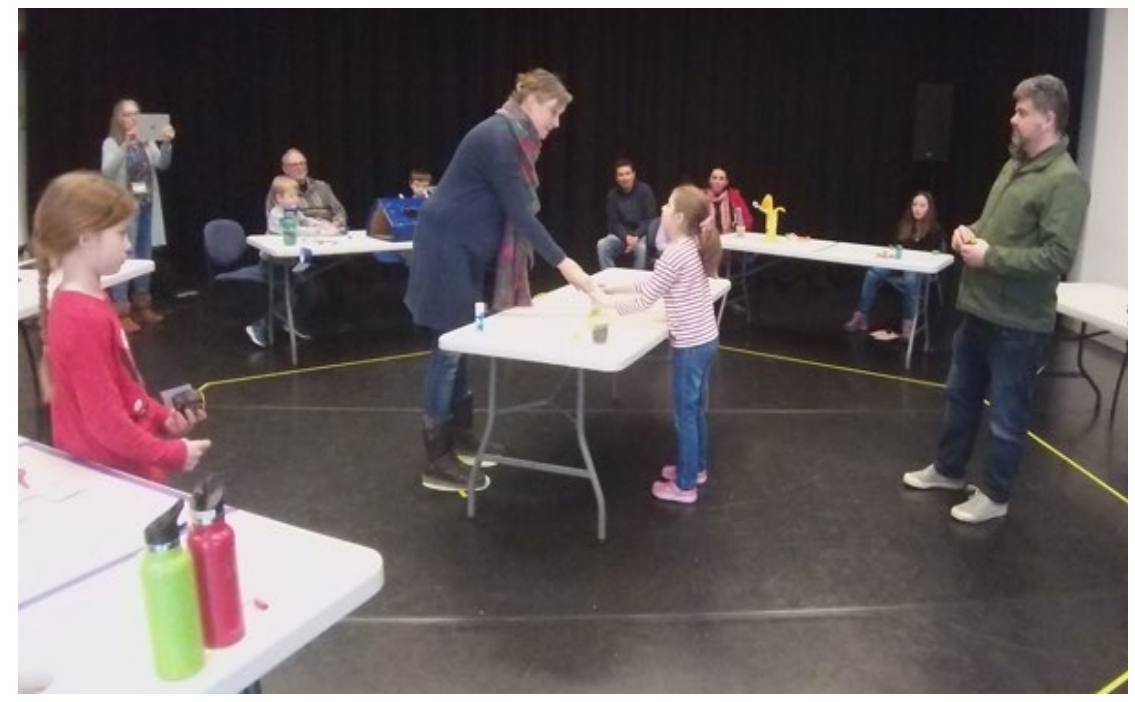

Figure 7. (a) Prototyping materials; (b) a family presenting their model in the hexagonal space during Workshop 2.

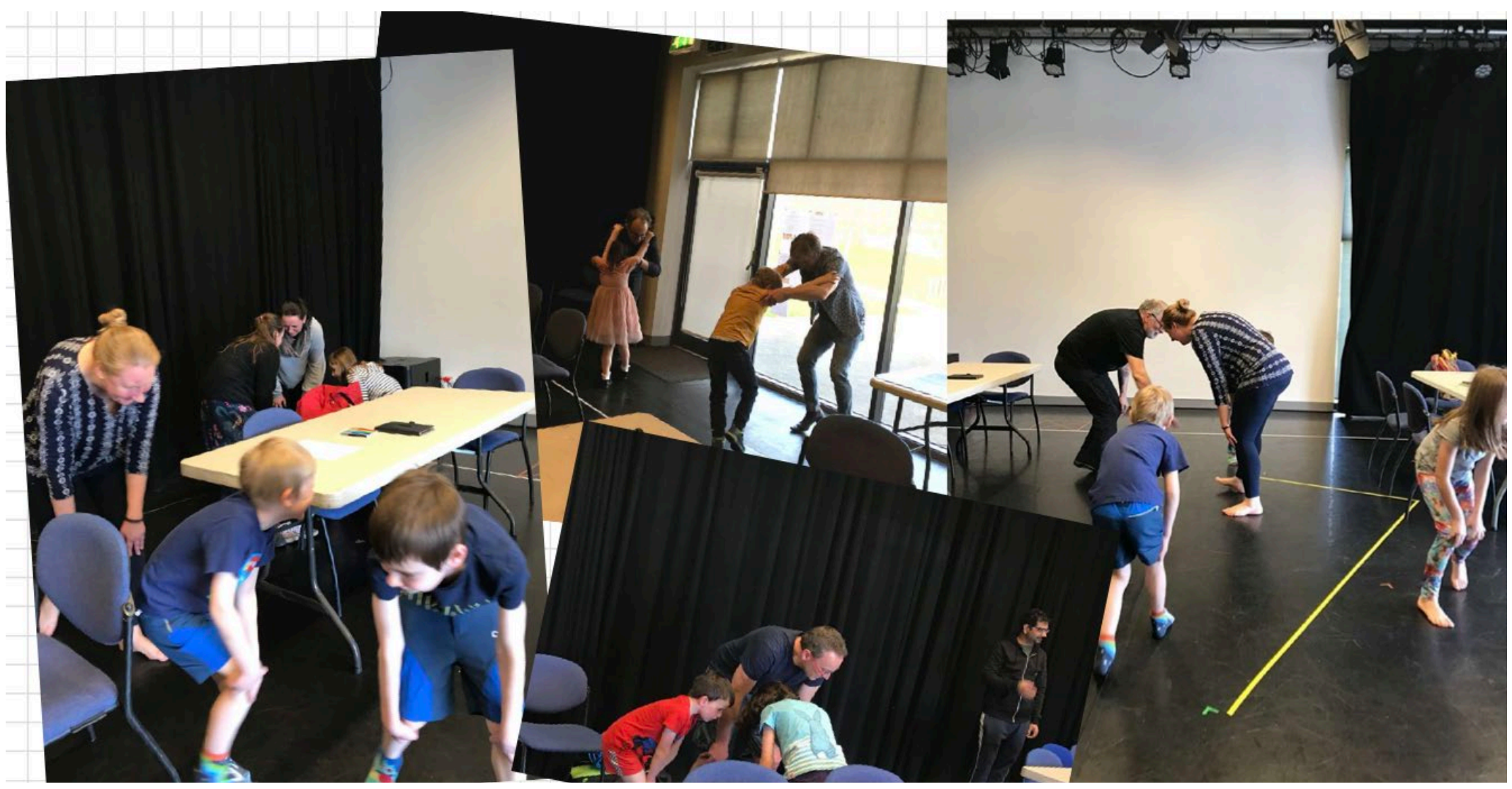

Figure 8. Irish Duels ice-breaker game

## 5 DISCUSSION

In this short paper, we do not provide an exhaustive account of the co-design data. Instead, we introduce two co-design orientations that emerged across the process and that we consider most relevant to family co-design around complex environmental issues: rebalancing child–adult power through playful, theatre-inspired status shifts, and carrying embodied forms of participation from workshop activity into the public mixed-reality event. Together, these orientations show how theatrical co-design can operate beyond artefact generation, shaping and rehearsing the roles, rhythms, and authority relations through which families collectively imagine change.

### 5.1 Rebalancing Child-Adult Power Relationships

Even when initial ideation is balanced, adults often dominate elaboration and evaluation unless specific interventions are made [30]. Because collaborative tasks could easily revert to adult-led guidance, we employed status-rebalancing mechanisms to create space for greater child initiative, a dynamic that can unsettle power relations [26].

Viewing power relations as enacted and relational [8], we found that theatre offered "as-if" spaces where authority could be redistributed by rehearsing alternative roles and establishing new habits of participation through playful practice. The imagination scanner exemplified this: a device that "measured" imagination and always ranked children above adults. Its humour enabled children to act as "imagination leaders," while adults adopted more receptive roles. Adults themselves later reflected that the scanner reminded them not to constrain the children and helped everyone enter a more creative play mode. For practitioners, a key takeaway is that playful status reversals, such as casting children as "imagination leaders," can redistribute authority and invite fuller contributions from less dominant voices.

### 5.2 Re-staging Workshop Dynamics

HCI has long explored embodied methods for enriching ideation, including bodystorming, soma design, tangible interaction, and physical play, all of which show how sensorimotor engagement can surface design possibilities [22, 9, 1, 15]. The key point here is that the workshops did not only generate design ideas; they also shaped how the final installation organised participation dramaturgically. Physical activities, tangible interactions, and the imagination scanner rehearsed roles and ways of participation that were later carried into *HOME: Zero*. The imagination scanner remained as the opening interaction, positioning children as imaginative leaders; adults were invited into a more playful, supportive role; families worked through physical props and shared decisions; and the performer/facilitator stayed inside the installation as a live guide. In this sense, the installation re-staged not only workshop artefacts, but also the participatory roles and relations through which those artefacts had been created.

### 5.3 Limitations and Future Work

This study is based on practice-based reflection across workshop plans, artefacts, facilitator accounts, and researcher positionality. As a result, claims about shifts in child–adult participation should be read as design insights rather than systematically measured behavioural outcomes. The compressed commission timescale also limited the trust-building and iteration usually associated with applied theatre and family co-design. Future work should examine these orientations through mixed-methods evaluations of participation, including interaction coding and analysis of material engagement, and through longer-term studies that allow theatrical co-design processes to develop across varied cultural and institutional settings.

## 6 CONCLUSION

This study positions applied theatre not as a toolkit of engagement exercises, but as a way of organising participatory design from workshop practice into public mixed-reality experience. In *HOME: Zero*, family co-design for environmental sustainability involved not only generating ideas, but also negotiating who could lead, who could play, and how families could act together in imagining domestic change. The two orientations discussed above show how theatrical co-design can rebalance child–adult participation through playful status shifts and re-stage selected workshop roles and ways of working within the final public event. These methods are not easily detachable techniques: they require contextual sensitivity, time, and skilled facilitation. For HCI, this points toward closer collaboration between design researchers and theatre practitioners in participatory projects involving families, embodiment, environmental imagination, and social change.

## ACKNOWLEDGMENTS

This work was supported by the Engineering and Physical Sciences Research Council [grant number EP/T022493/1] and the Horizon Centre for Doctoral Training at the University of Nottingham. HOME: Zero was created by Makers of Imaginary Worlds in partnership with Lakeside Arts, Sam Redway, and the Mixed Reality Lab, University of Nottingham. The project was commissioned by Nesta and National Gallery X, a partnership between the National Gallery and King's College London, supported by LEADD:NG, and co-designed with Nottingham families. The 2024–2025 UK national tour was supported by Arts Council England.